\documentclass[%
 reprint,
superscriptaddress,
 amsmath,amssymb,
 aps,
pra]{revtex4-2}

\usepackage{amsmath}
\usepackage[dvipsnames]{xcolor}
\usepackage{graphicx}  % Include figure files
\usepackage{physics}
\usepackage{dcolumn}  % Align table columns on decimal point
\usepackage{bm}  % Bold math
\usepackage{comment}

\usepackage{natbib}  % Bibliography management

\usepackage{hyperref}  % Should be loaded last to avoid conflicts
\usepackage{cleveref}  % Cross-referencing package (after hyperref)

\begin{document}

\preprint{APS/123-QED}

\title{Randomized benchmarking with non-Markovian noise and realistic finite-time gates}
%\title{Randomized benchmarking with realistic gate implementations and non-Markovian noise}
% Force line breaks with \\
% \thanks{A footnote to the article title}%

% Authors in a single line
\author{Antoine Brillant}
\affiliation{Pritzker School of Molecular Engineering, University of Chicago, Chicago, IL, USA}

\author{Peter Groszkowski}
\affiliation{National Center for Computational Sciences, Oak Ridge National Laboratory, Oak Ridge, TN, USA}

\author{Alireza Seif}
\affiliation{IBM Quantum, IBM T.J. Watson Research Center, Yorktown Heights, NY, USA}

\author{Jens Koch}
\affiliation{Department of Physics and Astronomy, Northwestern University, Evanston, Illinois 60208, USA}

\author{Aashish A. Clerk}
\affiliation{Pritzker School of Molecular Engineering, University of Chicago, Chicago, IL, USA}

% \collaboration{}%\noaffiliation

% \author{}
% \homepage{}
% \affiliation{
%  Second institution and/or address\\
%  This line break forced% with \\
% }%
% \affiliation{
%  Third institution, the second for Charlie Author
% }%
% \author{Delta Author}
% \affiliation{%
%  Authors' institution and/or address\\
%  This line break forced with \textbackslash\textbackslash
% }%

% \collaboration{}%
% \noaffiliation

\date{\today}% It is always \today, today,
             %  but any date may be explicitly specified

\begin{abstract}
% The rich interplay between non-Markovian noise and coherent drives gives rise to many phenomena. 
We analyze the impact of non-Markovian classical noise on single-qubit randomized benchmarking experiments, in a manner that explicitly models the realization of each gate via realistic finite-duration pulses.  Our new framework exploits the random nature of each gate sequence to derive expressions for the full survival probability decay curve which are non-perturbative in the noise strength. In the presence of non-Markovian noise, our approach shows that the decay curve can exhibit a strong dependence on the gate implementation method, with regimes of both exponential and power law decays.  We discuss how these effects can complicate the interpretation of a randomized-benchmarking experiment, but also how to leverage them to probe non-Markovianity. 
%We study the effect of this interplay on randomized benchmarking experiments by modeling the gates as realistic finite-duration pulses. We develop a framework which relies on the random nature of the gate sequences to derive expressions for the survival probability decay curve which are non-perturbative in the noise strength. Our model shows that in the presence of non-Markovian noise, the decay curve exhibits a strong dependence on the way that the gates are implemented, ranging from exponential to power law decay. We discuss how these effects can complicate the interpretation of the decay curve but also how to leverage them to probe non-Markovianty.
% \begin{description}
% \item[Usage]
% Secondary publications and information retrieval purposes.
% \item[Structure]
% You may use the \texttt{description} environment to structure your % % abstract;
% use the optional argument of the \verb+\item+ command to give the category % of each item. 
% \end{description}
\end{abstract}

%\keywords{Suggested keywords}%Use showkeys class option if keyword
                              %display desired
\maketitle

%\tableofcontents

\textit{Introduction---} Randomized benchmarking (RB) protocols are a powerful tool for characterizing errors in quantum processors. They rely on on the application of random gate sequences to robustly extract average properties of the noise without having to perform full quantum process tomography. 
%(something which typically scales exponentially with qubit number). 
For gate-independent Markovian error models, standard RB protocols predict an exponential decay of the survival probability \cite{knillRandomizedBenchmarkingQuantum2008a, magesanScalableRobustRandomized2011, magesanCharacterizingQuantumGates2012, helsenGeneralFrameworkRandomized2022,hashimPracticalIntroductionBenchmarking2024}. 
Under standard assumptions, the measured decay rate can be directly used to extract the average gate infidelity.
Unfortunately, in many relevant settings the dominant noise is non-Markovian (i.e.~correlated in time) \cite{kimErrorMitigationStabilized2024, weiCharacterizingNonMarkovianOffresonant2024, paladino1NoiseImplications2014a}. While RB protocols could still be useful in this context, it is not clear what they measure in the presence of non-Markovian noise.   RB protocols have been used in many situations where the Markovian assumption is not valid \cite{veldhorstAddressableQuantumDot2014, ryanRandomizedBenchmarkingSingle2009}.

The above concerns have motivated many recent works studying RB and non-Markovian noise \cite{figueroa-romeroGeneralFrameworkRandomized2022}. With such noise, it has been shown that the decay of the survival probability can be non-exponential \cite{fongRandomizedBenchmarkingCorrelated2017, figueroa-romeroRandomizedBenchmarkingNonMarkovian2021, fogartyNonexponentialFidelityDecay2015}, can converge more slowly to its mean \cite{ballEffectNoiseCorrelations2016, edmundsDynamicallyCorrectedGates2020} and can even be used to learn the noise spectrum \cite{zhangRandomisedBenchmarkingCharacterizing2023a}. However, typical approaches make approximations that can miss important physics.  In particular, the assumption of instantaneous gates can miss the potentially rich interplay between finite-duration gates and temporally-correlated noise. 

\begin{figure}
    \centering
    \includegraphics[width=1\linewidth]{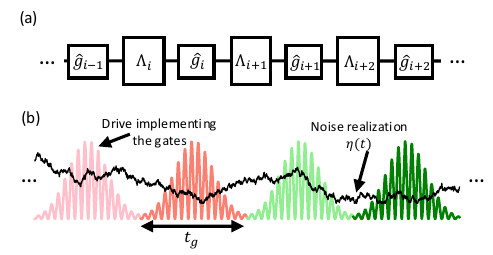}
    \caption{(a) A common approach used to model randomized benchmarking experiments. Each gate $\hat g_i$ is applied instantaneously, followed by the application of a noisy map $\Lambda_i$. The errors can be correlated between each time step, but the interplay with the drive implementing the gates would be hard to capture as one would need to obtain the corresponding $\Lambda_i$. (b) Artist's representation of the model analyzed in this letter. The noise is represented as a stochastic random variable $\eta(t)$ and the gates are implemented via finite time duration pulses.}
    \label{fig:cartoon}
\end{figure}

Here, we address these concerns by modeling single-qubit RB in a physically-motivated manner.  We consider a qubit that is driven by both finite-duration pulses (used to implement the chosen sequence of random gates) as well as classical non-Markovian, Gaussian noise (see e.g.~Fig.~\ref{fig:cartoon}.(b)). As we will show, this approach naturally captures the effects of noise correlations between adjacent gates which are present due to the finite-duration gate implementation. This effect would be hard to capture using other approaches (see e.g.~Fig.~\ref{fig:cartoon}.(a)). We develop a novel method that builds on Ref.~\cite{groszkowskiSimpleMasterEquations2023,vankampenCumulantExpansionStochastic1974, vankampenCumulantExpansionStochastic1974a} to understand the average dynamics of the qubit, averaged both over noise realizations and over gate sequences.  By exploiting the randomness of the control pulses, we are able to obtain expressions for the full survival probability decay curve that are non-perturbative in the noise strength.  
%This method  readily allows us to understand the interplay between the gate implementations and the noise by making use of the random nature of the gate sequences to obtain expressions for the survival probability decay curve that are non-perturbative in the noise strength. 
We show that the fully-averaged qubit dynamics is described by a time-dependent depolarizing channel, with the form of the time-dependent rate encoding the complex interplay of non-Markovian noise 
and the specific finite-time gate implementation.
%We find that even with non-Markovian noise and finite-duration gates, the decay of the survival probability is described by a time-dependent depolarizing channel (with the time dependence reflecting details of both the noise and the gates). 

We find that depending on noise parameters and gate implementation, non-Markovian noise can yield RB survival probabilities that decay either exponentially or as a power-law.  Perhaps not surprisingly, noise with a short correlation time (e.g. comparable to the gate time) yields exponential decay.  More surprisingly, the corresponding decay rate does not correspond to the average gate infidelity and can vary by a factor of almost 2 by changing the gate implementation. This raises questions about the interpretation of RB decay curves, even in the seemingly simple case where one obtains a purely exponential decay. Similar questions were raised in Ref.~\cite{proctorWhatRandomizedBenchmarking2017} due to the gauge freedom present in the definition of the average gate infidelity. In contrast, for our problem we show that this difference is instead linked to 
%physical mechanisms originating from 
the non-Markovian nature of the noise.

%Our model shows that in the presence of temporally correlated noise, the decay of the survival probability exhibits a strong dependence on the noise model, going from exponential for noise with a short correlation time to power law for noise with a long correlation time. Surprisingly, even when the noise has a short but finite correlation time, of the order of a few gate times, the decay is almost indistinguishable from being exponential. The decay rate however does not correspond to the average gate infidelity and can vary by a factor of almost 2 by changing the gate implementation. This makes the meaning of the RB decay rate questionable, even in the case where an exponential decay is observed.

\textit{Model---} We consider a standard single-qubit randomized benchmarking experiment \cite{knillRandomizedBenchmarkingQuantum2008a} in which each gate sequence consists of $L+1$ gates sampled uniformly from the Clifford group.  The qubit is subject to classical, Gaussian, stationary noise $\eta(t)$ (autocorrelation function $S(t)$), which couples (without loss of generality) to the qubit operator $\hat \sigma_z$. While our approach applies to any noise spectrum, in the main text we take as a paradigmatic example noise with an exponentially-decaying $S(t)$ (i.e.~an Ornstein–Uhlenbeck process),  with strength $\sigma$ and correlation time $\tau_c$.  Using a bar to denote noise averages:
\begin{align}
S(t-t') &\equiv \overline{ \eta(t) \eta(t')} = \sigma^2 \exp(-\lvert t - t' \rvert / \tau_c).
\end{align}
This form interpolates between Markovian and strongly non-Markovian limits, and is also directly relevant to many experiments (see e.g.~\cite{serniakHotNonequilibriumQuasiparticles2018}).  More general noise spectra are considered in the Supplemental Material (SM) \cite{supp}.   
%We also stress that our formalism applies to any choice of noise correlation function, as shown in the appendix. 
Each gate in the RB sequence is implemented by a pulse of duration $t_g$. Working in a rotating frame at the qubit frequency, the Hamiltonian for a single realization of the noise $\eta(t)$ and a specific gate sequence $\vec \beta$ (an $L+1$ length vector) is:
\begin{align}
\tilde H_{\eta}(t, \vec \beta) &= \vec \Omega(t, \vec \beta) \cdot \vec{\hat{\sigma}} + \eta(t) \hat \sigma_z , \label{hamiltonian}
% \\
% S(t-t') &\equiv \overline{ \eta(t) \eta(t')} = \sigma^2 \exp(-\lvert t - t' \rvert / \tau_c) ,
\end{align}
where $\vec \Omega(t,\vec \beta)$ encodes the pulse envelope implementing the sequence and $\vec{\hat{\sigma}}$ is the Pauli vector. 
%The overline denotes the average over the noise realizations. 
Importantly, because the noise is non-Markovian, the effect of the noise on the qubit depends on $\vec \Omega (t, \vec \beta)$ even though the control Hamiltonian is independent of $\eta(t)$. As the goal of this paper is to understand the global decay features of the survival probability, we neglect state preparation and measurement errors (which ideally do not impact the decay rate). We also assume that the zeroth gate of the sequence, $\beta_0$, is implemented instantaneously. As shown in the SM \cite{supp}, this has a negligible influence on our findings, but greatly simplifies our analysis.

We work in an interaction picture with respect to the noise-free Hamiltonian $\hat H_0(t,\vec \beta) = \vec \Omega(t, \vec \beta) \cdot \vec{\hat \sigma} $, where the dynamics is described by the stochastic Hamiltonian $\hat H_{\eta}(t, \vec \beta)= \eta(t) \hat \sigma_z(t, \vec \beta)$ with $\hat \sigma_z(t, \vec \beta)= \hat U_0^{\dag}(t,\vec \beta) \hat \sigma_z \hat U_0(t,\vec \beta)$ and $\hat U_0(t,\vec \beta)= \mathcal T \exp(-i\int_0^t dt' \hat H_0(t',\vec \beta))$.
%where $\mathcal T$ is the time ordering operator. 
This is a useful frame as there is no noise-free dynamics, letting us isolate the impact of noise.  Qubit evolution here is $\check{\mathcal U}_{\eta}(t,\vec \beta) \hat \rho \equiv \hat U_{\eta}(t,\vec \beta) \hat \rho \hat U^{\dag}_{\eta}(t,\vec \beta)$ where $\hat \rho$ is the initial qubit density matrix and $\hat U_{\eta}(t,\vec \beta) = \mathcal T \exp(-i\int_0^t dt' \hat H_{\eta}(t,\vec \beta))$. The quantity of interest in an RB experiment is the average survival probability:
\begin{equation}
    P_0(t) = \langle \overline{tr[\hat \rho_0 \hat \rho_{\eta}(t, \vec \beta) ]}\rangle_{\vec \beta}, \quad \hat \rho_{\eta}(t, \vec \beta) \equiv \check{\mathcal U}_{\eta}(t,\vec \beta) \hat \rho_0,
\end{equation}
where $\langle \boldsymbol{\cdot} \rangle_{\vec \beta} = \frac{1}{24^{L+1}} \sum_{\vec \beta} [\boldsymbol{\cdot}]$ is the average over the gate sequences. We assume that the initial state $\hat \rho_0$ is uncorrelated with the noise. 
%We emphasize that in this frame, $\hat \rho_\eta(t,\vec \beta)$ is equal to $\hat \rho_0$ in the absence of noise. 
Our task then reduces to computing the noise and sequenced averaged evolution superoperator:
\begin{equation}
    \check{\mathcal U}_{\text{avg}}(t) = \langle \overline{\check{\mathcal U}_{\eta}(t,\vec \beta)} \rangle_{\vec \beta}. \label{averaged_prop}
\end{equation}
This yields the survival probability via: $P_0(t) = tr(\hat \rho_0 \check{\mathcal U}_{\text{avg}}(t)\hat \rho_0)$. 

The most obvious next step is to compute the average over the noise in Eq.~\eqref{averaged_prop}. This is non-trivial:
even though $\eta(t)$ is Gaussian, the non-commuting structure of Eq.~\eqref{hamiltonian} gives rise to an infinite set of cumulants.  In some cases this hierarchy can be truncated to yield useful descriptions \cite{groszkowskiSimpleMasterEquations2023}, but this is an approach that is perturbative in the noise strength and that fails for noise with long correlation times.  
%even if the noise is Gaussian because the two terms in Eq.~\eqref{hamiltonian} do not commute. This gives rise to a non-trivial cumulant structure, which can be used to approximately describe dynamics as described in Ref.~\cite{groszkowskiSimpleMasterEquations2023}. However, this approach is perturbative in the noise strength and fails to accurately capture the dynamics for noise with long correlation times in regimes needed for RB.
We instead follow a different route, and first average over the random variable $\vec \beta$ (i.e.~over different random gate sequences) for a fixed noise realization $\eta(t)$. This yields an alternate kind of cumulant expansion, which we truncate to second order to get an approximation that is {\it non-perturbative} in the noise strength. As we show in the SM \cite{supp}, this alternate kind of cumulant approximation is similar to the one introduced in Refs.~\cite{foxCritiqueGeneralizedCumulant1976, kuboGeneralizedCumulantExpansion1962}, but is not restricted to noise with limited temporal correlations.  Instead, it is able to capture the impact of temporal correlations between non-adjacent gates to any order in the noise strength. 
%\AC{Hmm, will this be confusing, as in the expression eq 7, we only have correlations between adjacent gates.} \Antoine{I agree that it can look like it, but the rate in eq.7 contains correlations between distant gates too as it is inside the exponential in eq.6.. So the rate at time $t$ directly depends only on the noise in adjacent gates. But the rates at any well separated times are also correlated.   I added a comment after eq.7-8}
It is motivated by the decoupling properties of random gate sequences \cite{violaRandomDecouplingSchemes2005a, winickConceptsConditionsError2022}, which  generate an effective correlation-time of $t_g$ for the dynamics, making higher cumulants vanish on that time-scale. 

After averaging over gate sequences within this approximation, we obtain a propagator $\check{\mathcal U}_{\eta}(t)$ that still depends on the specific noise realization. After $m$ full gates are applied, it is given by:
\begin{gather}
    \check{\mathcal U}_{\eta}(m t_g) = \check{\mathcal I} + \Lambda_{\eta}(m t_g) \sum_{\alpha} \check{\mathcal D}[\hat \sigma_{\alpha}], \label{gate_avg_prop}\\
    \Lambda_{\eta}(m t_g) = \frac{1}{4} - \frac{1}{4} \exp(- 4 \int_0^{m t_g} dt' \Gamma_{\eta}(t')), \label {Lambda}
\end{gather}
where $\check{\mathcal I}$ is the identity superoperator and $\check{\mathcal D}$ is the Lindblad dissipator: $\check{\mathcal D}[\hat O]\hat \rho = \hat O \hat \rho \hat O^{\dag} - \frac{1}{2}(\hat O^{\dag} \hat O \hat \rho + \hat \rho \hat O^{\dag} \hat O)$. This is a depolarizing channel whose strength has a non-trivial dependence on $m$ as determined by the effective stochastic rate $\Gamma_\eta(t)$.  This in turn is given by (see SM \cite{supp}):

\begin{gather}
    \Gamma_{\eta}( nt_g + \tau) = 
        \frac{1}{3} \int_{(n-1)t_g}^{n t_g + \tau} dt' \eta(nt_g + \tau) \eta(t') f(nt_g + \tau,t') , \label{gamma} \\
    f(t_1,t_2) =\langle tr[\hat \sigma_z(t_1,\vec \beta) \hat \sigma_z(t_2,\vec \beta)] \rangle_{\vec \beta} ,\label{f_function}
\end{gather}
where $n \in \mathbb{Z}$ and $\tau \in [0,t_g)$ are defined by $t = \tau + nt_g$. Eq.~\eqref{gamma} is only valid for $n>0$; when $n=0$, the lower bound of the integral becomes $0$ 
\footnote{This is a consequence of our assumption that the initial state of the qubit is uncorrelated with the noise.}. 
%See SM [XXX} for a detailed derivation of Eqs.~\eqref{gate_avg_prop}-\eqref{Lambda} is provided in the supplement.  
%Here, we discuss the consequences of this result. 
The function $f(t_1,t_2)$ is the average overlap of the evolved noise operator at different times and is the only quantity that depends on the gate implementation. 
Note that for each noise realization, the sequence-averaged evolution is a time-dependent depolarizing channel. Further, the instantaneous rate $\Gamma_{\eta}(t)$ depends both on the behavior of the noise during the ``current" gate period, and during the previous gate.  The noise at even earlier times does not contribute directly to the instantaneous stochastic rate (as in this case, the earlier time and $t$ are separated by one or more complete random gates and hence a full twirl). However, longer-range temporal correlations will ultimately contribute once we perform an average over $\eta(t)$. 
%it does contribute to the dynamics as Eq.~\eqref{gamma} depends on $\eta(t)$ which can exhibit correlations between any two times. %This is a consequence of the finite-duration gates which do not apply a full twirl between any two consecutive gates. 

\begin{comment}
One can easily verify that it is $t_g$-periodic and that it vanishes whenever $t_1$ and $t_2$ are times that belong to two non-consecutive gates. This means that to lowest order in the noise, correlations only survive between adjacent gates.
    
\end{comment}

While Eq.~\eqref{gate_avg_prop} still has to be averaged over $\eta(t)$,
this remaining noise average has been greatly simplified: we now just need to average a single scalar quantity 
$\Lambda_{\eta}(t)$. Further, since $\Lambda_{\eta}(t)$ is the exponential of the square of the Gaussian random variable $\eta(t)$, computing its average reduces to computing a functional determinant; this can be done using several different methods (see e.g.~\cite{zhangSpectralEffectsDispersive2015}). We can also gain analytic insight by using approximate methods to average over the noise. In the following sections we show two simple approximate ways of computing the noise average. The first relies on a weak noise approximation, which remains valid at all times provided the correlation time of the noise is sufficiently small. The second, assumes that the noise is constant on the time scale of $t_g$, which is valid provided the correlation time of the noise is sufficiently long. In the SM \cite{supp}, we show that for any correlation time, one of these approximations is valid, meaning that together, they are enough to describe the decay of the survival probability in almost any regime.

Before diving into these approximations, we can compute the noise average exactly in two opposite limits \cite{supp}: Markovian and quasistatic noise. In the first case, $S(t) = \gamma \delta(t)$ and, as expected, the survival probability decays exponentially:
\begin{equation}
    P_0^{\text{M}}(m t_g)=\frac{1}{2}+\frac{1}{2}\exp \left(-\frac{4}{3}\gamma mt_g \right).
\end{equation}
In the second case, $S(t) = \sigma^2$ and the survival probability decays like a power law: 
\begin{equation}
    P_0^{\text{qs}}(m t_g) = \frac{1}{2} + \frac{1}{2\sqrt{1+ \frac{8}{3}\sigma^2 t_g^2 (m F_{\text{curr}} + (m-1) F_{\text{prev}})}},
\end{equation}
where $F_{\text{prev,curr}}$ are the integrals of $f(t_1,t_2)$ corresponding to the average overlap of the noise operator between times belonging to the same gate or adjacent gates respectively:
\begin{align}
     F_{\text{curr}} &= \frac{1}{t_g^2} \int_{0}^{t_g} dt_1 \int_0^{t_1}dt_2 f(t_1,t_2) ,\label{F_curr}\\
    F_{\text{prev}} &= \frac{1}{t_g^2} \int_{t_g}^{2t_g}dt_1 \int_0^{t_g} dt_2 f(t_1,t_2) .\label{F_prev}
\end{align}
These factors depend on how the gates are implemented, and will be useful going forward. 
% Importantly they only depend on the gate implementation.

\textit{Master equation description---} To average Eq.~\eqref{gate_avg_prop} over the noise, the simplest approximation is to perform a weak noise expansion and derive a time-local master equation \cite{groszkowskiSimpleMasterEquations2023}. This can be justified even in the long-time limit provided that the noise has a small correlation time, such that $\sigma^2 \tau_c t_g \ll 1$. %Starting from Eq.~\eqref{gate_avg_prop}, 
Making this approximation to second order in the noise, the equation of motion for the density matrix averaged over the noise and gate sequences $\hat \rho_{\text{avg}}(t) = \langle \overline{\hat \rho_{\eta}(t,\vec \beta)} \rangle_{\vec \beta}$ is:
\begin{equation}
    \partial_t \hat{\rho}_{\text{avg}}^{(2)}(t)=\overline{\Gamma_{\eta}(t)} \sum_{\alpha \in \{x,y,z\}} \mathcal D[\hat \sigma_{\alpha}] \hat \rho_{\text{avg}}(t) ,\label{2nd order PLME}
\end{equation}
where $\overline{\Gamma_{\eta}(t)}$ is the noise-averaged decay rate.  This rate is $t_g$ periodic after the first full gate is applied. We can then compute the survival probability:
\begin{align}
    P^{(2)}_0(mt_g) &= \frac{1}{2} + \frac{1}{2} 
    \exp(-4 \epsilon')
    \exp(-4 \epsilon (m-1) ) , \label{2nd order P0}\\
    \epsilon &= \int_{j t_g}^{(j+1) t_g} dt' \overline{\Gamma_{\eta}(t')} \quad \text{ for } j>0 ,\label{epsilon}
\end{align}
where Eq.~\eqref{epsilon} does not depend on $i$ due to the periodicity of $\overline{\Gamma_{\eta}(t)}$ and where $\epsilon' = \int_{0}^{t_g} dt' \overline{\Gamma_{\eta}(t')}$.
%is defined in the same way but for the first time step.  
We see that to second order in the noise strength, the survival probability decays exponentially. Further, the corresponding decay rate depends both on the gate implementation via $f(t_1,t_2)$ and on the noise correlation function $S(t_1-t_2)$.

\begin{figure}
    \centering
    \includegraphics[width=1\linewidth]{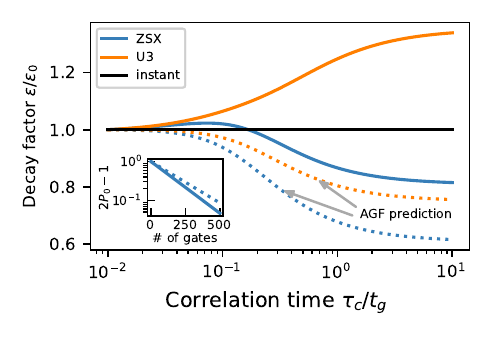}

    \caption{Survival probability decay factors $\epsilon/\epsilon_0$ from Eq.~\eqref{epsilon}  for three different gate implementations. $\epsilon_0=8\gamma_0/3$ is the value of $\epsilon$ in the limit $\tau_c/t_g \rightarrow 0$. The solid lines show the decay factor that would be observed in a RB experiment. The dotted lines show the decay factor $\epsilon'$ that would be predicted by calculating the average gate infidelity $\overline F$. These results are different due to noise correlations between adjacent gates. The inset shows the decay of the survival probability for RB experiment (solid line) and the AGF prediction (dotted line) with $\tau_c = 0.5 t_g$ and $\gamma_0 = 2.5 \times 10^{-3}$. All the values of $\tau_c/t_g$ on the $x$-axis of the main plot are within the regime of validity of Eq.~\eqref{2nd order P0}.}
    \label{fig:1}
\end{figure}

To study the dependence on choice of gate implementation, we consider three possible strategies for constructing pulses.  The first, $``ZSX"$, uses finite duration $X$ pulses interleaved by instantaneous $Z$ pulses.  A given Clifford gate is implemented as  $R_Z(\phi + \pi) \sqrt{X} R_Z(\theta + \pi) \sqrt{X} R_Z(\lambda)$ for appropriate $\phi, \theta, \lambda$.  This implementation is common in RB experiments, as all Clifford gates naturally take the same time (each $\sqrt{X}$ takes a time $t_g/2$). The second, $``U3"$, corresponds to the application of the constant pulse of duration $t_g$ that applies the shortest rotation that implements the gate on the Bloch sphere. The third, $``instant"$, corresponds to an instantaneous application of the gate followed by an idle time $t_g$. While this is unphysical, we include it as it is often used when modeling RB experiments, see e.g.~\cite{ballEffectNoiseCorrelations2016,  epsteinInvestigatingLimitsRandomized2014, figueroa-romeroGeneralFrameworkRandomized2022, figueroa-romeroRandomizedBenchmarkingNonMarkovian2021, fongRandomizedBenchmarkingCorrelated2017}. 

Fig.~\ref{fig:1} plots survival probability decay rates $\epsilon$ predicted by Eq.~\eqref{epsilon} as a function of noise correlation time 
$\tau_c$. 
To meaningfully compare the impact of varying  
$\tau_c$, we also vary $\sigma$ with $\tau_c$ so that the variance $\gamma_0$ of the random phase acquired during a time $t_g$ remains fixed, $\gamma_0 = \int_0^{t_g} dt_1 \int_0^{t_1} dt_2 S(t_1-t_2)$. 
% Thus, the amount of undriven dephasing during a time $t_g$ is fixed. 
%This is also motivated by the fact that $S(t)$ approximates the Markovian limit when $\tau_c/t_g \rightarrow 0$ while it corresponds to quasistatic (DC) noise when $\tau_c/t_g \rightarrow \infty$. 
As shown in the SM \cite{supp}, all the values of $\tau_c/t_g$ shown on the $x$-axis of Fig.~\ref{fig:1} are within the regime of validity of Eq.~\eqref{2nd order P0} for values of $\gamma_0 = 2.5 \times 10^{-3}$. We see that $\epsilon$ varies by a factor of almost $2$ between the $``ZSX"$ and $``U3"$ gate implementations when $\tau_c \sim 10 t_g$. This dependence on choice of gate implementation
can be interpreted as arising from the ability of finite-duration gate pulses to perform a kind of dynamical decoupling cancellation or enhancement of correlated noise.  As such, the implementation dependence is absent in the extreme Markovian limit.  Note however that even for small but non-zero correlation times (e.g.~$\tau_c \approx 0.1 t_g$), the error rate per gate is higher than in the true Markovian limit $\tau_c/t_g \rightarrow 0$. We see that even a small departure from the Markovian limit can have an impact.  

It is natural to ask whether the RB decay rate $\epsilon$ that we find is related to the average gate fidelity (AGF), defined as $\overline{F} = \int d\psi \bra{\psi} \check{\mathcal U}_{\text{avg}}(t_g)[\dyad{\psi}]\ket{\psi}$. This definition is equivalent to the survival probability after one gate $P_0(t_g)$, meaning that $\overline{F} = (\exp(-4\epsilon')+1)/2$.
Using standard RB theory for Markovian noise \cite{magesanScalableRobustRandomized2011},
the estimated AGF extracted from the decay curve $\check{\overline{F}}$ would be determined from $\check{\overline{F}} = (\exp(-4 \epsilon)+1)/2$.
As shown in Fig.~\ref{fig:1}, $\epsilon$ (plain lines) and $\epsilon'$ (dotted lines) significantly differ from each other meaning that 
the AGF does not accurately predict the RB decay curve, even if it is exponential. 
%the average gate infidelity predicted by the AGF significantly differ from its true value, even if the decay looks exponential.
%\AC{Don't we have to write what the prediction for $\bar{F}$ would be based on $\epsilon$?  Can't take this as being obvious.} \Antoine{Yes ok, I changed the above a little bit.}
Since $\overline F = P_0(t_g)$ is a physical quantity, the difference between $\check{\bar F}$ and $\bar F$ is not due to our choice of gauge for the AGF (as was studied in Ref.~\cite{proctorWhatRandomizedBenchmarking2017}). Instead, 
it reflects the simple fact that with non-Markovian noise, errors at time $t$ will depend on what happened during earlier gates, something that cannot be captured by characterizing individual gates in isolation. 
%it is because in the presence of non-Markovian noise, the gates can't be characterized in isolation as their performance depends on the length of the sequence.

%\AC{The 2nd part of this paragraph needs to be re-written, too much repetition, and the key point is perhaps not clear enough.  Could emphasize the point that Luke made, that this result is not surprising, the AGF isn't a good way of characterizing errors when things are non-Markovian. } \Antoine{Ok I think it's clear now.}
%\AC{End of edits Dec 14}
%\Antoine{Ok, I modified this paragraph quite a bit, I think that it is clearer now.}

\textit{Coarse grained noise approximation---} 
For noise with longer correlation times, the straighfrorward perturbative approach to Eq.~\eqref{gate_avg_prop} is no longer valid, as terms that are higher-order in the noise cannot be neglected.
%Obtaining an accurate long time description of the survival probability curve via the PLME method can be quite cumbersome for noise with long correlation times as one has to go to higher orders in the noise strength. The problem lies in the fact that non-Markovian effects on timescales longer than $t_g$ are not captured in the lowest order master equation as the twirling of the noise suppresses correlations among gates separated by more than one gate time.
While one could attempt to use a partial resummation method, here we take a different approach. For cases where the noise has a long correlation time $\tau \gg t_g$, 
%To make progress in obtaining the full decay of the survival probability curve for noise with a long correlation time, 
we can coarse grain the noise with negligible induced error, i.e.~replace $\eta(t)$ in Eq.~\eqref{gamma} by a set of stochastic random variables $\theta_i/t_g \equiv \frac{1}{t_g} \int_{i t_g}^{(i+1)t_g}dt' \eta(t')$. 
%If $\tau_c \gtrapprox t_g$, then this approximation should be accurate. 
This replacement greatly simplifies the noise averaging, which reduces to the evaluation of a finite matrix determinant. Using this approximation (see SM \cite{supp} for details) we can express the survival probability after the application of $m$ gates in terms of
two $m \times m$ matrices: the correlation matrix of the coarse-grained noise  $\boldsymbol{\Sigma}$ and a tri-diagonal matrix $\textbf{F}$ which encodes the effect of the finite-time gate implementation.  Letting $\textbf{1}$ denote the identity matrix, we have:
\begin{align}
    P_0(m t_g) &= \frac{1}{2} + \frac{1}{2\sqrt{\textrm{det}( \textbf{1} + \frac{8}{3} \boldsymbol{\Sigma} \textbf{F} )}}, \label{surv_prob_constant_noise}\\
    \Sigma_{i,j} &= \overline{\theta_i \theta_j}, \label{corr_func}\\
    \textrm{F}_{i,j} &= F_{\text{curr}}\delta_{i,j} + \frac{1}{2} F_{\text{prev}} (\delta_{i,j+1} + \delta_{i,j-1}). \label{F_matrix}
\end{align}
 Note that in the simple limit where gates are implemented instantaneously, Eq.~(\ref{surv_prob_constant_noise}) reproduces  the expression derived in  Ref.~\cite{fongRandomizedBenchmarkingCorrelated2017}.
 It however goes beyond this, as it also lets us understand the impact of finite gate times and different choices of gate implementation. Indeed, since each of the matrix $\textbf{F}$ is Toeplitz (except for the first time step), it can be seen as a scalar renormalizing the strength of the correlations in $\boldsymbol{\Sigma}$. In the SM \cite{supp}, we show that for noise with long correlation times, an additional coarse-grained approximation makes this statement rigourous. The renormalization factor is then simply given by $F = F_{\text{curr}} + F_{\text{prev}}$.

 %To gain intuition on how this impacts the decay curve, we make a further approximation 
 %where the noise is constant on the time-scale of $2t_g$ : $\theta_i \approx \theta_{i+1}$, but can still vary on longer timescales. 
 %\AC{This is very confusing Antoine.  If the noise is constant between any pair of adjacent gates, then it is completely constant.  Please reword this more carefully.} \Antoine{Ok}
  %Then Eq.~\eqref{surv_prob_constant_noise} stays the same with $\textbf{F}_{i,j}= F_{\text{curr}}\delta_{i,j} + F_{\text{prev}} \delta_{i,j} (1-\delta_{i,0})$, which is proportional to $\textbf{1}$ after the first gate time. This means that for noise with $\tau_c \gg t_g$, the effect of the gate implementation is to renormalize the noise strength by the factor $F=F_{\text{curr}} + F_{\text{prev}}$.
  %\AC{Please be consistent with matrix subscripts.  No commas.} \Antoine{Ok, I left the commas because of the $i+1$ indices. Its consistent now.}

\begin{figure}
    \centering
    \includegraphics[width=1.\linewidth]{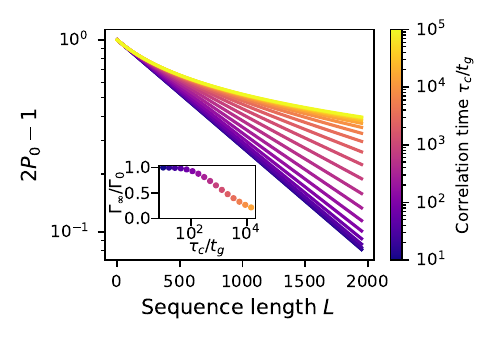}
    \caption{Decay of the survival probability as a function of the number of gates $L$ for various values of $\tau_c$ (colorbar). The gates are implemented via the '$ZSX$' gate decomposition. The decay follows a power law for $L t_g \ll \tau_c$ and an exponential decay with rate $\Gamma_{\infty}$ for $Lt_g \gg \tau_c$. The inset shows $\Gamma_{\infty}$ as a function of $\tau_c$, with $\Gamma_0$, the initial decay rate. $\Gamma_{\infty}$ is obtained for each curve by fitting its long time decay behavior to an exponential $A \exp(-\Gamma_{\infty} L t_g) + B$. The noise strength $\sigma$ is fixed in the same way as in Fig.~\ref{fig:1} with $\gamma_0 = 6.125 \times 10^{-4}$.} 
    \label{fig:2}
\end{figure}

In Fig.~\ref{fig:2} we used Equations \eqref{surv_prob_constant_noise}-\eqref{F_matrix} to study the survival probability for noise with long, but finite correlation times. We find that as a function of the sequence length $m$, the curve initially decays as a power law, before transitioning to exponential decay with rate $\Gamma_\infty$ for large $m$. We also see in the inset that $\Gamma_\infty$ decreases with $\tau_c$. This is expected since here $\tau_c \gg t_g$, the noise can remain correlated amongst many gates leading to randomized dynamical decoupling effects \cite{violaRandomDecouplingSchemes2005a}. This is in contrast with changing the gate implementation, which introduces noise averaging on the scale of $t_g$ and which could either increase or decrease the decay rate.

We can therefore isolate the main contributions coming from the gate implementation and the noise correlation function. The gate implementation changes the effective strength of the noise by introducing averaging effects on the timescale of $t_g$ while the correlation function of the noise changes the functional form of the noise by allowing averaging effects on the scale of multiple gate times.
%\Antoine{I reformulated the paragraph above (starting with In fig3...). This is basically repeating the same thing, but I wonder if it can be useful as it summarizes a little bit more information? than the previous par.}

Fig.~\ref{fig:2} also reveals that for correlation times on the order of $10t_g$, the entirety of the survival probability decay curve looks almost exponential. In these regimes it would therefore be hard to resolve the correlation time of the noise by looking solely at the survival probability decay curve. Nevertheless, seeing a variation in the decay rate when changing the gate implementation could be used as a flag for non-Markovianity.

\textit{Conclusion---} 
Using a generalized cumulant approximation that goes beyond simple perturbative approaches, 
we studied the interplay between finite-duration gates and non-Markovian classical noise in RB experiments. 
%Using a generalized cumulant approximation, we have been able to go beyond perturbative approaches in the noise correlation time to obtain the full survival probability decay curve. 
We find that the decay remains exponential in the presence of noise with correlation times on the order of $t_g$, but with a decay rate that is highly dependent on the physical gate implementation and that does not correspond to the average gate infidelity.  This raises multiple questions concerning the interpretation of RB experiments when noise is non-Markovian.  
%usefulness of RB as a protocol to obtain the average error per gate in the presence of non-Markovian noise. 
In future work, it would be interesting to employ our generalized cumulant expansion to study other protocols involving random gate sequences such as randomized compiling \cite{winickConceptsConditionsError2022, wallmanNoiseTailoringScalable2016}.

\textit{Acknowledgments}--- We thank Lorenza Viola for useful discussions, and Luke Govia for a careful reading of the manuscript and for helpful comments.
This research was sponsored by the Army Research Office and was accomplished under Grant No.~W911NF-23-1-0116.  
A.S. was sponsored by the Army Research Office under Grant Number W911NF-21-1-0002. The views and conclusions contained in this document are those of the authors and should not be interpreted as representing the official policies, either expressed or implied, of the Army Research Office or the U.S. Government. The U.S. Government is authorized to reproduce and distribute reprints for Government purposes notwithstanding any copyright notation herein.
This manuscript has been coauthored by UT-Battelle, LLC, under contract DE-AC05-00OR22725 with the US Department of Energy (DOE). The US government retains and the publisher, by accepting the article for publication, acknowledges that the US government retains a nonexclusive, paid-up, irrevocable, worldwide license to publish or reproduce the published form of this manuscript, or allow others to do so, for US government purposes. DOE will provide public access to these results of federally sponsored research in accordance with the DOE Public Access Plan (https://www.energy.gov/doe-public-access-plan).

\let\oldaddcontentsline\addcontentsline% Store \addcontentsline
\renewcommand{\addcontentsline}[3]{}% Make \addcontentsline a no-op
\bibliography{references}
\let\addcontentsline\oldaddcontentsline% Restore \addcontentsline

\clearpage

\newpage 
% \clearpage
\thispagestyle{empty}
\onecolumngrid
\begin{center}
\textbf{\large Supplemental Material: Randomized benchmarking with non-Markovian noise and realistic finite-time gates }
\end{center}

\begin{center}
Antoine Brillant${}^{1}$, Peter Groszkowski${}^{2}$, Alireza Seif${}^{3}$, Jens Koch${}^{4}$, Aashish A. Clerk${}^1$\\
\emph{
${}^1$Pritzker School of Molecular Engineering, University of Chicago, Chicago, IL 60637, USA \\
${}^2$National Center for Computational Sciences, Oak Ridge National Laboratory, Oak Ridge, TN, USA\\
${}^3$IBM Quantum, IBM T.J. Watson Research Center, Yorktown Heights, NY, USA\\
${}^4$Department of Physics and Astronomy, Northwestern University, Evanston, Illinois 60208, USA}\\
(Dated: \today)
\end{center}

% \author{Antoine Brillant}
% \affiliation{Pritzker School of Molecular Engineering, University of Chicago, Chicago, IL, USA}

% \author{Peter Groszkowski}
% \affiliation{National Center for Computational Sciences, Oak Ridge National Laboratory, Oak Ridge, TN, USA}

% \author{Alireza Seif}
% \affiliation{IBM Quantum, IBM T.J. Watson Research Center, Yorktown Heights, NY, USA}

% \author{Jens Koch}
% \affiliation{Department of Physics and Astronomy, Northwestern University, Evanston, Illinois 60208, USA}

% \author{Aashish Clerk}
% \affiliation{Pritzker School of Molecular Engineering, University of Chicago, Chicago, IL, USA}

% Prevent references from appearing in the TOC

\tableofcontents

\section{Derivation of the sequence average propagator }
 In this section we elaborate on the cumulant expansion in the variable $\vec \beta$ that is used to obtain the approximate sequence averaged propagator Eq.~\eqref{gate_avg_prop}. We first outline the derivation and then demonstrate its validity, which relies on the fact that higher cumulants of the quantity that we average over almost vanish.

\subsection{Derivation}

\subsubsection{Noise and sequence dependent propagator}
Our starting point is the total Hamiltonian:
\begin{equation}
    \tilde H_{\eta}(t, \vec \beta) = \vec \Omega(t, \vec \beta) \cdot \vec{\hat{\sigma}} + \eta(t) \hat A ,
\end{equation}
which is analogous to, but slightly more general than Eq.~\eqref{hamiltonian} in the main text since here, we consider that the noise couples to an arbitrary Hermitian operator $\hat A$ and the correlation function of the noise $S(t-t')= \overline{\eta(t) \eta(t')}$ is not specified. As in the main text, we consider that each gate takes a finite time $t_g$ to be applied and $\vec \Omega (t,\vec \beta)$ is the time dependent pulse sequence that applies the gate sequence indexed by $\vec \beta$. We denote the Clifford gates by the unitary operators $\hat g_j \in \mathcal G$ where $\mathcal G$ is the Clifford group. In particular $\hat g_{\beta_i}$ is the $i$th Clifford gate in the sequence.

Because we assume that there are no preparation and measurement errors, we can apply a perfect instantaneous gate $\hat g_{\beta_0}$ at the beginning of each sequence, which we refer to as the zeroth gate. This is equivalent to being able to perfectly prepare the qubit in any state, which is one of our assumptions. We also show numerically in Sec.~\ref{sec:perfect_first_gate} that this instantaneous gate has a negligible effect on the decay curve of the survival probability for the regimes considered in this paper. Since one of our assumptions is that there are no state preparation and measurement errors, we could think of this zeroth intstantaneous gate as perfectly preparing the qubit. This gate is not applied via finite-time pulses and therefore not contained in $\vec \Omega (t,\vec \beta)$, we therefore add it to the dynamics by hand.

The first step to obtain the average sequence propagator is to go in the interaction frame with respect to the noise free Hamiltonian $\hat H_0(t,\vec \beta) = \vec \Omega(t, \vec \beta) \cdot \vec{\hat \sigma}$. The unitary operator associated to this change of frame is given by: 
\begin{equation}
    \hat U_0(t,\vec \beta) \hat g_{\vec \beta_0} =\mathcal T \exp(-i\int_0^t dt' \hat H_0(t',\vec \beta)) \hat g_{\vec \beta_0},
\end{equation}    
where $\hat g_{\beta_0}$ is explicitly factored out as mentionned above and where $\mathcal T$ is the time ordering operator. In this frame the Hamiltonian is given by the time evolved noise operator $\hat A$:
\begin{align}
    \hat H_{\eta}(t,\vec \beta) &= \eta(t) \hat g_{\beta_0}^{\dag} \hat A(t, \vec \beta) \hat g_{\beta_0} \label{noisy_H},\\
    \hat A(t, \vec \beta) &= \hat U_0^{\dag}(t,\vec \beta) \hat A  \hat U_0(t,\vec \beta).
\end{align}
Given a time $t = n t_g + \tau$ where $n$ is the number of full Clifford gates (excluding the zeroth gate) that have been applied prior to time $t$ and $\tau \in [0,t_g)$, we can write:
\begin{equation}
    \hat U_0(n t_g+\tau,\vec \beta) = \hat V_{\beta_{n+1}}(\tau) \hat g_{\beta_n} \dots \hat g_{\beta_1} \label{U_0},
\end{equation}
where $\hat V_i(\tau)$ is the unitary operator implementing the $i$th Clifford gate, such that the following constraints:
\begin{equation}
    \hat V_i(0) = \hat 1, \quad \hat V_i(t_g) = \hat g_i,
\end{equation}
are respected and where $\hat 1$ is the identity operator.

The evolution for a single gate sequence and a single noise realization is therefore governed by the Liouville-von Neumann equation:
\begin{align}
    \dot{\hat \rho}_{\eta}(t,\vec \beta) = \eta(t) \check{\mathcal L}(t,\vec \beta) \hat \rho_{\eta}(t, \vec \beta), \\
   \check{\mathcal L}(t, \vec \beta)[\cdot] = -i [\hat g_{ \beta_0}^{\dag} \hat A(t, \vec \beta) \hat g_{\beta_0}, \cdot ] \label{mathcalL},
\end{align}
which we can formally solve to obtain:
\begin{align}
    \hat \rho_{\eta}(t,\vec \beta) &= \mathcal T \exp(\int_0^t dt' \eta(t') \check{\mathcal L}(t', \vec \beta) ) \hat \rho_0 \label{single_traj_evol}\\
    &\equiv \check{\mathcal U}_{\eta}(t, \vec \beta)\hat \rho_0,
\end{align}
where $\hat \rho_0$ is the initial state which we assume to be uncorrelated with the noise.

\subsubsection{Introduction of the superoperator cumulants}

As mentioned in the main text, we are interested in performing the average over the random variable $\vec \beta$ first. To this end, we employ the well known method of cumulant expansions as introduced in Ref.~\cite{kuboGeneralizedCumulantExpansion1962, foxCritiqueGeneralizedCumulant1976,vankampenCumulantExpansionStochastic1974,vankampenCumulantExpansionStochastic1974a}. Specifically, we will follow a very similar approach as was done in Ref.~\cite{foxCritiqueGeneralizedCumulant1976}, which requires us to define the $n$th cumulant superoperator:
\begin{equation}
    \check{\mathcal C}_k(\vec t) \equiv \langle \langle \check{\mathcal L}(t_{k}, \vec \beta) \dots \check{\mathcal L}(t_1, \vec \beta)\rangle \rangle_{\vec \beta} \\
    \equiv \langle \langle k,k-1, \dots ,1 \rangle \rangle_{\vec \beta},
\end{equation}
where $\vec t$ is the vector containing the times $t_{k},\dots, t_1$ which are in decreasing order from left to right such that $t_{\alpha+1}>t_{\alpha}$. The second equality in the previous equation is simply a way to lighten the notation by simply identifying the time indices of the cumulant. A formal definition of the cumulant average $\langle \langle \cdot \rangle \rangle_{\vec \beta}$ is given in~\cite{foxCritiqueGeneralizedCumulant1976}, but we emphasize that it corresponds to the connected part of the average of time-ordered products of $\check{\mathcal L}(t,\vec \beta)$. We also stress that the time $t_{\alpha}$ does not necessarily correspond to a time during which the $\alpha$th Clifford gate is applied. The first few cumulants (using the analogous notation $\langle k,k-1,\dots,1 \rangle_{\vec \beta}$ to denotes the moments) are given in terms of moments by:
\begin{align}
    \langle \langle 1 \rangle \rangle_{\vec \beta} &=  \langle 1 \rangle_{\vec \beta}\\
    \langle \langle 2,1 \rangle \rangle_{\vec \beta} &=  \langle 2,1\rangle_{\vec \beta} - \langle 2\rangle_{\vec \beta} \langle 1\rangle_{\vec \beta}\\
    \langle \langle 3,2,1 \rangle \rangle_{\vec \beta} &=  \langle 3,2,1\rangle_{\vec \beta} - \langle 3,2\rangle_{\vec \beta} \langle 1\rangle_{\vec \beta} - \langle 3,1\rangle_{\vec \beta} \langle 2\rangle_{\vec \beta} -  \langle 3\rangle_{\vec \beta} \langle 2,1\rangle_{\vec \beta} + \langle 3\rangle_{\vec \beta} \langle 2\rangle_{\vec \beta}  \langle 1\rangle_{\vec \beta} + \langle 3\rangle_{\vec \beta} \langle 1\rangle_{\vec \beta}  \langle 2\rangle_{\vec \beta} \label{cumulant_example}
\end{align}
We note that because the superoperator $\check{\mathcal L}(t,\vec \beta)$ does not commute with itself at different times, these superoperator cumulants are slightly different than the more common scalar cumulants, as defined in Ref.~\cite{kuboGeneralizedCumulantExpansion1962}.

We can evaluate the first two cumulants explicitely. To do so, we can make use of some well known single qubit Clifford group averaging identities (see e.g.~\cite{meleIntroductionHaarMeasure2024} for an in dept introduction). 
\begin{align}
   \frac{1}{24}\sum_{g\in \mathcal G}  \hat g^{\dag} \hat O \hat g &= \frac{1}{2} tr(\hat O) \hat 1\\
    \frac{1}{24}\sum_{g\in \mathcal G} \hat g^\dag \hat O_1 \hat g \hat \rho \hat g^\dag \hat O_2 \hat g &= \frac{1}{24}\bigg( \big(-4 tr(\hat O_1 \hat O_2) + 8 tr(\hat O_1)tr(\hat O_2)\big) \hat \rho + \big(8 tr(\hat O_1 \hat O_2) - 4 tr(\hat O_1) tr(\hat O_2) \big)\hat 1 \bigg),
\end{align}
where $\hat O_{1,2}$ are arbitrary operators and $\hat \rho$ is any density matrix. Using these, we compute the first and second cumulant by averaging over $\hat g_{\beta_0}$:
\begin{align}
    \check{\mathcal C}_1(t) &= \langle -i [\hat g^{\dag}_{\beta_0} \hat A(t,\vec \beta) \hat g_{\beta_0}, \cdot ] \rangle_{\vec \beta} = 0, \label{cumulant_1}\\
    \check{\mathcal C}_2(t_2, t_1) &= \bigg\langle - \big[\hat g^{\dag}_{\beta_0} \hat A(t_2,\vec \beta) \hat g_{\beta_0},[\hat g^{\dag}_{\beta_0} \hat A(t_1,\vec \beta) \hat g_{\beta_0}, \cdot ]\big] \bigg\rangle_{\vec \beta} = \frac{1}{3} tr \big(\hat A(t_2,\vec \beta) \hat A(t_1,\vec \beta) \big) \sum_{\alpha \in \{x,y,z \}} \check{\mathcal D} [\hat \sigma_{\alpha}], \label{cumulant_2}
\end{align}
where $\check{\mathcal D}[\hat O]\hat \rho = \hat O \hat \rho \hat O^{\dag} - \frac{1}{2}(\hat O^{\dag} \hat O \hat \rho + \hat \rho \hat O^{\dag} \hat O)$ is the Lindblad dissipator. We made use of Eq.~\eqref{cumulant_1} in Eq.~\eqref{cumulant_2} and we used the fact that $tr (\hat A(t,\vec \beta))=0$. Therefore the second cumulant is the first non-vanishing contribution to the cumulant expansion. We notice that when exponentiated, Eq.~\eqref{cumulant_2} yields a depolarizing channel. This is not surprising as applying a full Clifford twirl (here with $\hat g_{\beta_0}$) always tailors a noise channel to a depolarizing channel. One could in fact show that this holds for the cumulants of any order. This means that the superoperator cumulants commute with each other for all times.

We also stress, that while this result seems to be a consequence of the application of a first instantaneous gate, we would obtain a very similar result even without $\hat g_{\beta_0}$. Indeed, as soon as a full Clifford gate is applied, (i.e for $t>t_g$), we could average over $\hat g_{\beta_1}$ to obtain the same result. Therefore, even without the initial noiseless gate, each cumulant correspond to a depolarizing channel when $t_\alpha>t_g$ $\forall t_\alpha \in \vec t$. While we could in principle do this calculation, it leads to a complicated correction for $t<t_g$ and we argue that the effect on the decay of the survival probability is negligible as shown numerically in Sec.~\ref{sec:perfect_first_gate}.

\subsubsection{Cumulant expansion}

We can use the superoperator cumulants to compute the sequence average of $\check{\mathcal U}_{\eta}(t,\vec \beta)$. Formally, the sequence averaged propagator is given by the following cumulant expansion (see e.g.~, Ref~\cite{foxCritiqueGeneralizedCumulant1976}):
\begin{align}
    \langle \check{\mathcal U}_{\eta}(t, \vec \beta) \rangle_{\vec \beta} &= \exp \bigg\langle \bigg\langle \mathcal T \exp( \int_0^t dt' \eta(t') \check{\mathcal L}(t', \vec \beta))-\check{\mathcal I} \bigg\rangle \bigg\rangle_{\vec \beta} \label{cumulant_expansion_kubo}\\
    &= \exp\bigg( \sum_{k=1}^{\infty}\frac{1}{k!} \bigg\langle \bigg\langle \mathcal T \big(\int_0^t dt' \mathcal \eta(t') \check{\mathcal L}(t',\vec \beta) \big)^k \bigg\rangle \bigg\rangle_{\vec \beta} \bigg)\\
    &= \exp\bigg( \sum_{k=1}^{\infty} \int_0^t dt_{k-1} \int_0^{t_{k-1}} dt_{k-2} \dots \int_0^{t_1} dt_0 \mathcal \eta(t_{k-1})\eta(t_{k-2})\dots \eta(t_{0}) \check{\mathcal C}_{k}(\vec t) \bigg)  \label{cumulant_expansion}.
\end{align}
We note that Eq.~\eqref{cumulant_expansion_kubo} is in the form introduced by Kubo in Ref.~\cite{kuboGeneralizedCumulantExpansion1962}, but that Eq.~\eqref{cumulant_expansion} is the more common way of writing the cumulant expansion. We also stress that there is no time ordering prescription for the first exponential on each line because all the cumulants are depolarizing channels and therefore commute with each others.

Next, we truncate the cumulant expansion in Eq.~\eqref{cumulant_expansion} to second order. Similar truncations of cumulant expansions are often used to average over the noise. In this context, they are usually justified via some weak noise approximation or the fact that the noise has a small correlation time compared to the time scale of the evolution of the system (see e.g.~Refs. \cite{vankampenCumulantExpansionStochastic1974, vankampenCumulantExpansionStochastic1974a}). However, in our case the average is performed over the gate sequences $\vec \beta$ and does not rely on the correlation time of the noise being small. Instead, as we show in the next section, this truncation is justified by the fact that the application of random Clifford gates suppresses correlations between products of superoperators $\check{\mathcal L}(t,\vec \beta)$ separated by more than a full Clifford gate. Only keeping the second cumulant (and remembering that the first cumulant vanishes), we obtain the following approximation for the sequence averaged superoperator propagator:
\begin{align}
      \langle \check{\mathcal U}_{\eta}(t, \vec \beta)\rangle_{\vec \beta} &= \exp( \int_0^t dt_2 \int_0^{t_2 }dt_1 \eta(t_2) \eta(t_1)  \check{\mathcal C}_2(t_2,t_1) )\\
      &=  \exp( \frac{1}{3} \int_0^t dt_2 \int_0^{t_2 }dt_1 \eta(t_2) \eta(t_1) \big \langle tr \big(\hat A(t_2,\vec \beta) \hat A(t_1,\vec \beta) \big) \big \rangle_{\vec \beta}  \sum_{\alpha \in \{x,y,z \}} \check{\mathcal D} [\hat \sigma_{\alpha}])\\
      &= \check{\mathcal I}- \frac{1}{4} \bigg( 1-\exp(\frac{-4}{3} \int_0^t dt_2 \int_0^{t_2 }dt_1 \eta(t_2) \eta(t_1) \big \langle tr \big(\hat A(t_2,\vec \beta) \hat A(t_1,\vec \beta) \big) \big \rangle_{\vec \beta} ) \bigg) \sum_{\alpha \in \{x,y,z \}} \check{\mathcal D} [\hat \sigma_{\alpha}]\label{second_cumulant_prop}.
\end{align}
Upon replacing $\hat A$ by $\hat \sigma_z$, we recover Eqs.~\eqref{gate_avg_prop}-\eqref{f_function}. We also emphasize that a much simpler way to obtain this propagator would be to expand $\check{\mathcal U}_{\eta}(t, \vec \beta)$ up to second order in powers of $\eta(t)$, then perform the gate sequence average before re-exponentiating the result. While this is equivalent to our result, we argue that our approach that relies on neglecting higher cumulants is easier to justify.

\subsection{Justification of the second order cumulant approximation}

\begin{figure}[h]
    \centering
    \includegraphics[width=1\linewidth]{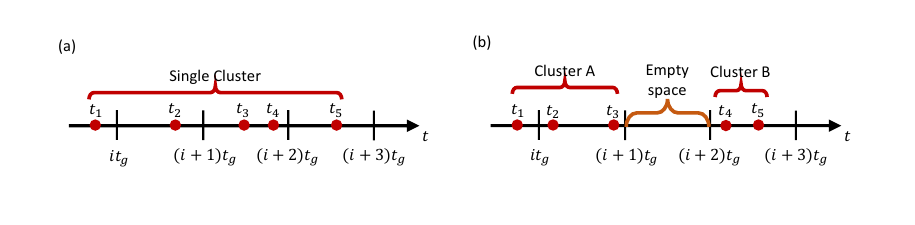}
    \caption{Cluster property of the cumulant expansion. (a) When all the times involved in a cumulant are clustered together (no consecutive times are separated by a full gate), the cumulant does not vanish. (b) When two or more consecutive times are separated by a full gate, the associated moment factorizes and the cumulant vanishes.}
    \label{fig:cumulant_sketch}
\end{figure}

To justify that the second order cumulant approximation is valid, we will show next that the higher cumulants $\check{\mathcal C}_{k>2}(\vec t)$ vanish whenever two consecutive times $t_{\alpha +1}, t_\alpha$ in $\vec t$ are separated by more than a full Clifford gate (see Fig~\ref{fig:cumulant_sketch}). This means that the contribution from higher cumulants is only non-zero when all the times $\vec t$ are clustered such that two consecutive times, $t_\alpha$ and $t_{\alpha +1}$ are in the same or adjacent gates. We will then show that of all the possible ways to arrange the times in $\vec t$, this condition is rarely fulfilled such that the contribution of all higher cumulant can be neglected compared to the contribution from the second order cumulant.

 We thus want to show that higher cumulants vanish when two consecutive times are separated by more than a full Clifford gate.

\subsubsection{Factorization of higher moments}

The first step is to show that the average of a product of superoperators $\langle k, k-1, \dots, 1\rangle_{\vec \beta}$ can be factored into the average of two smaller products $\langle k,\dots, \alpha+1 \rangle_{\vec \beta}\langle \alpha,\dots, 1 \rangle_{\vec \beta}$ whenever $t_{\alpha+1}$ and $t_{\alpha}$ are separated by more than one full Clifford gate. It is well known that for classical variables, this factorization property of the moments implies that the associated cumulant vanishes \cite{kuboGeneralizedCumulantExpansion1962}. This intuitively makes sense since a cumulant corresponds to a connected average and if a moment can be factorized, then the average is not connected. The generalization to operator valued random variables that do not commute is not trivial but was shown in Ref.~\cite{foxCritiqueGeneralizedCumulant1976} by Fox  in the case where the random variables follow the same time ordering prescription as in the present work.

We will first show this for a product of three superoperators. The generalizations to higher moments will then rely on the same argument. We make use the following identity:
\begin{equation}
    -i[\hat W^{\dag} \hat A \hat W, \cdot] = -i \hat W^{\dag} [\hat A, \hat W \cdot \hat W^{\dag}] \hat W,
\end{equation}
which is valid for any operator $\hat A$ and any unitary operator $\hat W$.  Using this, we can rewrite 
the superoperator $ \check{\mathcal L}(t, \vec \beta)$ (c.f.~Eq.~\eqref{mathcalL})
%:
% \begin{equation}
%     \check{\mathcal L}(t, \vec \beta)[\cdot] = -i [\hat g_{ \beta_0}^{\dag} \hat A(t, \vec \beta) \hat g_{\beta_0}, \cdot ],
% \end{equation}
in such a way that factors out the contribution from the full Clifford gates that have been applied up to time $t$. To this end, we decompose $t=nt_g + \tau$, (equivalently $t_{\alpha}=n_{\alpha}t_g + \tau_{\alpha}$) in the number $n$ ($n_\alpha$) of full gates applied at time $t$ ($t_\alpha$) (excluding the instantaneous $\hat g_{\beta_0}$) and the time $\tau (\tau_\alpha) \in [0,t_g)$ in the gate being applied at time $t$ ($t_\alpha$). We also define the superoperators corresponding to the applications of the $i$th Clifford gate and the partial application of the $i$th Clifford gate up to time $\tau$ :

\begin{align}
    \check{\mathcal G}_i[\cdot] &= \hat g_i [\cdot] \hat g_i^{\dag}\\
    \check{\mathcal V}_i(\tau)[\cdot] &= \hat V_i(\tau) [\cdot] \hat V_i^{\dag}(\tau).
\end{align} 
Finally, we introduce the following superoperator which will allow us to easily factor out the contribution of $\hat g_{\beta_0}$:
\begin{equation}
    \check{\mathcal L}_0 = -i[\hat A, \cdot].
\end{equation}
Using these, we can write the propagator of the noise free dynamics as:
\begin{equation}
    \check{\mathcal U}_0(\tau_{\alpha} + n_{\alpha}t_g,\vec \beta) = \check{\mathcal V}_{\vec \beta_{n_\alpha+1}}(\tau) \check{\mathcal G}_{\vec \beta_{n_\alpha}} \dots \check{\mathcal G}_{\vec \beta_{0}} \label{prop_factorization}.
\end{equation}
This then allows us to write Eq.~\eqref{mathcalL} as:
\begin{equation}
    \check{\mathcal L}(t_\alpha,\vec \beta)=\check{\mathcal U}_0^{-1}(t_\alpha,\vec \beta) \check{\mathcal L}_0 \check{\mathcal U}_0(t_\alpha,\vec \beta) \label{factorization1}
\end{equation}

Next we want to show that if two times are separated by more than one full gate (i.e., $n_{\alpha+1}>n_\alpha + 1$), then the average of the product factorizes in products of smaller averages. Again, we consider times that are ordered so that $t_\alpha>t_{\alpha-1}$. First, we consider the product of three superoperators and assume that $n_3>n_2+1$. Using the factorization \eqref{factorization1}, we obtain:
\begin{multline}
    \langle \check{\mathcal L}(t_3,\vec \beta)  \check{\mathcal L}(t_2,\vec \beta) \check{\mathcal L}(t_1,\vec \beta)\rangle_{\vec \beta} =  \big \langle  \check{\mathcal U}_0^{-1}(t_3,\vec \beta) \check{\mathcal L}_0 \check{\mathcal U}_0(t_3,\vec \beta) \check{\mathcal U}_0^{-1}(t_2,\vec \beta) \check{\mathcal L}_0 \check{\mathcal U}_0(t_2,\vec \beta) \check{\mathcal U}_0^{-1}(t_1,\vec \beta) \check{\mathcal L}_0 \check{\mathcal U}_0(t_1,\vec \beta)
    \big \rangle_{\vec \beta}\\
    =  \big \langle  \check{\mathcal G}_{\beta_0}^{-1}\dots  \underline{\check{\mathcal G}_{\beta_{n_3}}^{-1}} \check{\mathcal V}_{\beta_{n_3+1}}^{-1}(\tau_3) \check{\mathcal L}_0 \check{\mathcal V}_{\beta_{n_3+1}}(\tau_3) \underline{\check{\mathcal G}_{\beta_{n_3}}} \dots \check{\mathcal G}_{\beta_{n_2+1}} \check{\mathcal V}_{\beta_{n_2+1}}^{-1}(\tau_2)   \check{\mathcal L}_0 \check{\mathcal U}_0(t_2,\vec \beta) \check{\mathcal U}_0^{-1}(t_1,\vec \beta) \check{\mathcal L}_0 \check{\mathcal U}_0(t_1,\vec \beta) \big \rangle_{\vec \beta}, \label{three_prod_avg}
\end{multline}
where in the second line, we simply expanded the propagators $\check{\mathcal U}$ according to Eq.~\eqref{prop_factorization} and simplified the superoperators $\check{\mathcal G}$ that cancel each other between $\beta_{n_3}$ and $\beta_{n_2}$. We stress that because $n_3>n_2+1$, $\check{\mathcal G}_{\beta_{n_3}}$ does not cancel with $\check{\mathcal G}_{\beta_{n_2 + 1}}$. This means that $\check{\mathcal G}_{\beta_{n_3}}$ appears twice (underlined) in Eq.~\eqref{three_prod_avg}, both before and after the first $\check{\mathcal L}_0$. On the other hand, it does not appear anywhere else in this equation since all times to the right of the second $\check{\mathcal G}_{\beta_{n_3}}$ are smaller than $t_3$. This allows us to compute the average over $\check{\mathcal G}_{\beta_{n_3}}$ separately. Furthermore, since this average is over Clifford gates, and because the Clifford group is a unitary-$2$ design, this average has to commute with any unitary superoperator (see  e.g.~\cite{meleIntroductionHaarMeasure2024}). Doing this average, we obtain:
\begin{multline}
    \langle \check{\mathcal L}(t_3,\vec \beta)  \check{\mathcal L}(t_2,\vec \beta) \check{\mathcal L}(t_1,\vec \beta)\rangle_{\vec \beta}
    \\=  \big \langle  \check{\mathcal G}_{\beta_0}^{-1}\dots  \underline{\big \langle \check{\mathcal G}_{\beta_{n_3}}^{-1} \check{\mathcal V}_{\beta_{n_3+1}}^{-1}(\tau_3) \check{\mathcal L}_0 \check{\mathcal V}_{\beta_{n_3+1}}(\tau_3) \check{\mathcal G}_{\beta_{n_3}} \big \rangle_{\beta_{n_3}}} \dots \check{\mathcal G}_{\beta_{n_2+1}} \check{\mathcal V}_{\beta_{n_2+1}}^{-1}(\tau_2)   \check{\mathcal L}_0 \check{\mathcal U}_0(t_2,\vec \beta) \check{\mathcal U}_0^{-1}(t_1,\vec \beta) \check{\mathcal L}_0 \check{\mathcal U}_0(t_1,\vec \beta) \big \rangle_{\vec \beta}\\
    = \underline{\big \langle \check{\mathcal G}_{\beta_{n_3}}^{-1} \check{\mathcal V}_{\beta_{n_3+1}}^{-1}(\tau_3) \check{\mathcal L}_0 \check{\mathcal V}_{\beta_{n_3+1}}(\tau_3) \check{\mathcal G}_{\beta_{n_3}} \big \rangle_{\vec \beta}}  \big \langle  \check{\mathcal G}_{\beta_0}^{-1}\dots   \dots \check{\mathcal G}_{\beta_{n_2+1}} \check{\mathcal V}_{\beta_{n_2+1}}^{-1}(\tau_2)   \check{\mathcal L}_0 \check{\mathcal U}_0(t_2,\vec \beta) \check{\mathcal U}_0^{-1}(t_1,\vec \beta) \check{\mathcal L}_0 \check{\mathcal U}_0(t_1,\vec \beta) \big \rangle_{\vec \beta}, \label{factor_1}
\end{multline}

where we used the fact that the average over $\beta_{n_3}$ (underlined) commutes with everything between the second and third lines. In the last line, we changed the average on the first factor to be over $\vec \beta$, which is equivalent since it only contains $\beta_{n_3}$ and $\beta_{n_{3}+1}$. Finally, the first term commutes with any unitary superoperator, meaning that we can multiply is by $\check{\mathcal G}_{i}^{-1}$ and $\check{\mathcal G}_{i}$ on the left and on the right respectively without changing its value:
\begin{multline}
    \big \langle \check{\mathcal G}_{\beta_{n_3}}^{-1} \check{\mathcal V}_{\beta_{n_3+1}}^{-1}(\tau_3) \check{\mathcal L}_0 \check{\mathcal V}_{\beta_{n_3+1}}(\tau_3) \check{\mathcal G}_{\beta_{n_3}} \big \rangle_{\vec \beta} = \check{\mathcal G}_{\beta_{n_0}}^{-1} \dots \check{\mathcal G}_{\beta_{n_3 -1}}^{-1} \big \langle \check{\mathcal G}_{\beta_{n_3}}^{-1} \check{\mathcal V}_{\beta_{n_3+1}}^{-1}(\tau_3) \check{\mathcal L}_0 \check{\mathcal V}_{\beta_{n_3+1}}(\tau_3) \check{\mathcal G}_{\beta_{n_3}} \big \rangle_{\vec \beta} \check{\mathcal G}_{\beta_{n_3 -1}} \dots \check{\mathcal G}_{\beta_0}\\
    = \big \langle \check{\mathcal L}(t_3,\vec \beta) \big \rangle_{\vec \beta}. \label{double_avg}
\end{multline}
 Between the two lines, we averaged over $\vec \beta$, which does not affect the result since the left hand side was already averaged over $\vec \beta$. Using Eqs.~\eqref{factor_1}-~\eqref{double_avg} together, we obtain:
 \begin{equation}
     \langle \check{\mathcal L}(t_3,\vec \beta)  \check{\mathcal L}(t_2,\vec \beta) \check{\mathcal L}(t_1,\vec \beta)\rangle_{\vec \beta} = \langle \check{\mathcal L}(t_3,\vec \beta) \rangle_{\vec \beta} \langle  \check{\mathcal L}(t_2,\vec \beta) \check{\mathcal L}(t_1,\vec \beta)\rangle_{\vec \beta} 
 \end{equation}

This shows that when $n_3>n_2+1$, the product of the three point average factorizes in the product of a one point average and a two point average.

To generalize this result, we could redo the same steps for a moment of $k$ factors. We would then see that whenever $t_{\alpha}$ and $t_{\alpha+1}$ are separated by more than a full Clifford gate, the superoperator $\check{\mathcal G}_{\alpha+1}$ will appear exactly twice in the equation equivalent to Eq.~\eqref{three_prod_avg}, allowing us to average over it separately. The general result is therefore:
\begin{equation}
    \langle \check{\mathcal L}(t_k,\vec \beta) \dots  \check{\mathcal L}(t_1,\vec \beta) \rangle_{\vec \beta} = \langle \check{\mathcal L}(t_k,\vec \beta) \dots  \check{\mathcal L}(t_{\alpha+1},\vec \beta) \rangle_{\vec \beta}\langle \check{\mathcal L}(t_\alpha,\vec \beta) \dots  \check{\mathcal L}(t_1,\vec \beta) \rangle_{\vec \beta} \quad \text{ if } n_{\alpha+1} > n_\alpha + 1. \label{factorization}
\end{equation}
Or equivalently: $\langle k, \dots, 1\rangle_{\vec \beta} = \langle k,\dots \alpha+1 \rangle_{\vec \beta} \langle \alpha,\dots 1 \rangle_{\vec \beta}$.

We have shown that the moments factor when two times $t_\alpha, t_{\alpha +1}$ are separated by at least one full Clifford gate. Together with the argument proposed in Ref.~\cite{foxCritiqueGeneralizedCumulant1976}, this means that the higher cumulants vanish whenever this condition is respected.

As an example, we can look at the fourth cumulant. After using the fact that the first moment vanishes, we obtain:
\begin{align}
    \langle \langle 4,3,2,1 \rangle \rangle_{\vec \beta} &= \langle 4,3,2,1 \rangle_{\vec \beta}
    -
    \langle
    4,3
    \rangle_{\vec \beta}
    \langle
    2,1
    \rangle_{\vec \beta}
    -
    \langle
    4,2
    \rangle_{\vec \beta}
    \langle
    3,1
    \rangle_{\vec \beta}
    -
    \langle
    4,1
    \rangle_{\vec \beta}
    \langle
    3,2
    \rangle_{\vec \beta}.
\end{align}
We see that if for example $t_3$ and $t_2$ are separated by more than a Clifford gate, the first term on the right hand side will factorize such that it cancels the second term. The third term vanishes because if $n_3>n_2 +1$ then $n_4>n_2+1$, meaning that $\langle 4,2\rangle_{\vec \beta}= \langle 4\rangle_{\vec \beta} \langle 2 \rangle_{\vec \beta}$, which is equal to 0 because the first moment vanishes. The fourth term vanishes using the same argument.

The fact that higher cumulants vanish when two consecutive times are separated by more than a full Clifford gate means that the domain of integration of each term in the exponential in Eq.~\eqref{cumulant_expansion}:
\begin{align}
    \langle \check{\mathcal U}_{\eta}(t, \vec \beta) \rangle_{\vec \beta} = \exp\bigg( \sum_{k=1}^{\infty} \int_0^t dt_{k-1} \int_0^{t_{k-1}} dt_{k-2} \dots \int_0^{t_1} dt_0 \mathcal \eta(t_{k-1})\eta(t_{k-2})\dots \eta(t_{0}) \check{\mathcal C}_{k}(\vec t) \bigg)
\end{align}
can be reduced to one where all the times are close to $t_{k-1}$. Each integral (except for the first one) will therefore have a domain of integration of at most $2t_g$, comprising times where $t_{\alpha-1}$ is in the same gate as $t_{\alpha}$ or in the adjacent gate. Assuming that the noise $\eta(t)$ scales with the strength parameter $\eta_0$, this means that the $k$th order term in the sum will scale like $\eta_0^k (2t_g)^{k-1} t$. When expanding the exponential in the previous equation in powers of $\eta_0$, the dominant contribution to any order will then come from the second order cumulant. Therefore, the second order cumulant expansion should be valid at all times, provided $\eta_{\text{max}} t_g \ll 1$.

One can notice that in the case of noise with small correlation times and fixing the noise strength as in Fig.~\ref{fig:2} in the main text, $\eta_0$ could be in general too big for the previous argument to hold. In this case, the validity of the approximation relies on the fact that, for noise with a short correlation time, a more standard cumulant expansion (in powers of the noise) is valid. As we show in the next section, Eq.~\eqref{gate_avg_prop} recovers this more conventional cumulant expansion in the case of noise with weak correlation times. An easy way to verify this is to expand Eq.~\eqref{gate_avg_prop} to second order in the noise, to compute the noise average and to re-exponentiate it. The obtained result is the same as obtained in the next section.

\section{Second order PLME}

In this section we show how Eq.~\eqref{2nd order PLME} and \eqref{2nd order P0} in the main text were derive from Eq.~\eqref{second_cumulant_prop}. As shown in Ref.~\cite{groszkowskiSimpleMasterEquations2023}, a time-local master equation can be derived from the superoperator propagator $\check{\mathcal U}(t)$ of the dynamics via:
\begin{equation}
    \dot{\hat \rho}(t) = \dot{\check{\mathcal U}}(t) \check{\mathcal U}^{-1}(t) \hat \rho(t).
\end{equation}
We can then expand $\check{\mathcal U}(t)$ in powers of the noise to obtain an approximate master equation that is valid for noise with small correlation times. To second order in the noise strength, this effective master equation is:
\begin{equation}
    \dot{\hat \rho}^{(2)}(t) = \dot{\check{\mathcal U}}^{(2)}(t) \hat \rho^{(2)}(t),
\end{equation}
where $\check{\mathcal U}^{(2)}(t)$ is the second order term in the weak noise expansion of $\check{\mathcal U}(t)$. Using, Eq.~\eqref{second_cumulant_prop}, we obtain that for our RB experiment:
\begin{align}
    \dot{\check{\mathcal U}}^{(2)}(t) &= \overline{\dot{\Lambda}_{\eta}(t)} \sum_{\alpha} \check{\mathcal D}[\hat \sigma_{\alpha}]\\
    &= \overline{\Gamma_{\eta}(t)} \sum_{\alpha} \check{\mathcal D}[\hat \sigma_{\alpha}]\\
    &= \Gamma(t) \sum_{\alpha} \check{\mathcal D}[\hat \sigma_{\alpha}],
\end{align}
where we used the quantities defined in Eqs.~\eqref{Lambda}-\eqref{gamma} the main text:
\begin{align}
    \Lambda_{\eta}(t) &= \frac{1}{4} - \frac{1}{4} \exp(- 4 \int_0^{t} dt' \Gamma_{\eta}(t')) ,\\
    \Gamma_{\eta}(\tau + nt_g) &= 
        \frac{1}{3} \int_{(n-1)t_g}^{n t_g + \tau} dt' \eta(nt_g + \tau) \eta(t') f(nt_g + \tau,t'). 
\end{align}

This recovers Eq.~\eqref{2nd order PLME} from the main text. Then to obtain the survival probability, we have:
\begin{equation}
    P_0^{(2)} = tr(\hat \rho_0 \hat \rho^{(2)}(t)).
\end{equation}
To obtain $\hat \rho^{(2)}(t)$ we can simply solve the master equation, which is easy because $\dot{\check{\mathcal U}}^{(2)}(t)$ commutes which itself at all times. We obtain:
\begin{align}
    \hat \rho^{(2)}(t) &= \exp(\int_{0}^{t} dt' \Gamma(t') \sum_{\alpha}[\hat \sigma_\alpha]) \hat \rho_0 \\
    &= \bigg(\mathcal I + \frac{1}{4} \bigg(1- \exp \big(-4 \int_0^{t} dt' \Gamma(t') \big) \bigg) \sum_{\alpha}[\hat \sigma_\alpha] \bigg)\hat \rho_0,
\end{align}
which yields the survival probability:
\begin{align}
    P_0^{(2)} &= 1 -  \frac{1}{2} \bigg(1- \exp \big(-4 \int_0^{t} dt' \Gamma(t') \big) \bigg)\\
    &= \frac{1}{2} + \exp \big(-4 \int_0^{t} dt' \Gamma(t') \big). \label{surv_prob_2nd_supp}
\end{align}
This is the result from the main text.

We stress that starting the derivation from Eq.~$\eqref{single_traj_evol}$ instead of Eq.~\eqref{second_cumulant_prop} we would have obtained the same master equation to second order. To do so, we would have averaged over the noise realizations and the gate sequences simultaneously in the second order expansion. This highlights the fact that the result \eqref{surv_prob_2nd_supp} is equivalent but does not depend on the approximation of the previous section. Instead it depends on the correlation time of the noise being small compared to the total evolution time.

\section{Coarse-grained noise approximation}

In this section we derive Eqs.~\eqref{surv_prob_constant_noise}-\eqref{F_matrix},  an approximation that is valid provided the value of the noise $\eta(t)$ varies weakly during a single gate. Our starting point is Eq.~\eqref{cumulant_expansion}, in which we can replace $\eta(t)$ by its average value during the $i$th time step of length $t_g$: $\theta_i/t_g = \frac{1}{t_g}\int_{i t_g}^{(i+1)t_g}dt'\eta(t')$. Doing this, we obtain:
\begin{equation}
    \Lambda_{\eta}(mt_g) = \frac{1}{4} -\frac{1}{4}\exp(-\frac{4}{3} (F_{\text{curr}} \theta_0^2 + \sum_{i=1}^{m-1} F_{\text{curr}} \theta_i^2 +  F_{\text{prev}} \theta_i \theta_{i-1} )), \label{Lambda_const_noise}
\end{equation}
where we need to understand the subscript $\eta$ on $\Lambda_{\eta}(t)$ as meaning that this is still dependent on a particular trajectory and we defined:
\begin{align}
    F_{\text{curr}} &= \frac{1}{t_g^2} \int_{0}^{t_g} dt_1 \int_0^{t_1}dt_2 f(t_1,t_2),\\
    F_{\text{prev}} &= \frac{1}{t_g^2} \int_{t_g}^{2t_g}dt_1 \int_0^{t_g} dt_2 f(t_1,t_2).
\end{align}
We highlight that the limits of the integrals are respectively times in the same gate or adjacent gates, hence the names $\text{curr}$ and $\text{prev}$. The benefit of this is that instead of averaging over all the possible trajectories, we only need to average over the vector $\vec \theta = \{ \theta_i \}_{i=0, \dots, m-1}$, which is Gaussian distributed with covariance matrix:
\begin{equation}
    \boldsymbol{\Sigma}_{i,j} \equiv \overline{\theta_i \theta_j} = \int_{it_g}^{(i+1)t_g} dt_1 \int_{j t_g}^{(j+1)t_g} dt_2 S(t_1-t_2),
\end{equation}
We can then rewrite Eq.~\eqref{Lambda_const_noise} as:
\begin{equation}
    \Lambda_{\eta}(mt_g) =\frac{1}{4}-\frac{1}{4}\exp(-\frac{4}{3} \vec{\theta}^{\intercal} \cdot \textbf{F} \cdot \vec \theta),
\end{equation}
where the symmetric positive definite matrix $\textbf{F}$ of size $m \times  m$ is defined via:
\begin{equation}
    \textbf{F}_{i,j}= F_{\text{curr}}\delta_{i,j} + \frac{1}{2} F_{\text{prev}} (\delta_{i,j+1} + \delta_{i,j-1}).
\end{equation}
We can then average over the noise:
\begin{equation}
    \overline{\Lambda_{\eta}(m t_g)} =\frac{1}{4}-\frac{1}{4 I_0} \int d \vec \theta \exp(-\frac{1}{2}\vec \theta^{\intercal} \boldsymbol{\Sigma}^{-1} \vec \theta) \exp( -\frac{4}{3} \vec \theta^{\intercal} \textbf{F} \vec \theta),
\end{equation}
where $I_0(mt_g) = \int d \vec \theta \exp(-\frac{1}{2}\vec \theta^{\intercal} \boldsymbol{\Sigma}^{-1} \vec \theta)$ is the normalization constant. Since this integral is Gaussian, we can evaluate it exactly:
\begin{equation}
    \overline{\Lambda_{\eta}(m t_g)} = \frac{1}{4} - \frac{1}{4} \frac{1}{\sqrt{\lvert \textbf{1} + \frac{8}{3} \boldsymbol{\Sigma} \textbf{F}  \rvert}},
\end{equation}
where $\textbf{1}$ is the identity matrix and $\lvert \cdot \rvert$ denotes the determinant. This leads to the survival probability:
\begin{equation}
    P_0 =   \frac{1}{2}+ \frac{1}{2} \frac{1}{\sqrt{\lvert \textbf{1} +\frac{8}{3}\boldsymbol{\Sigma} \textbf{F}\rvert}}.
\end{equation}
We can also make a further simplifying assumption by treating the noise as constant between any two consecutive steps. In this case, we replace $\textbf{F}$ by:
\begin{equation}
    \textbf{F}_{i,j}= F_{\text{curr}}\delta_{i,j} + F_{\text{prev}} \delta_{i,j} (1-\delta_{i,0}),
\end{equation}
and we see that appart from the first time step, $\textbf{F}$ is proportional to the identity matrix. We can therefore think of the effect of the gate implementations as a renormalization of the gate time, in the limit where the correlation time of the noise is much greater than $t_g$.

\subsection{Validity of the coarse-grained noise approximation}

To verify the validity of the constant noise approximation, we can simply look at how far the random variable $\eta(t)$ is from its average during a gate time. The easiest way to do this is to look at the variance of the difference of $\eta(i t_g + \tau) - \theta_i$ with $\tau \in [0,1)$:
\begin{align}
    \xi(\tau) &= \frac{\overline{(\eta(it_g + \tau) - \theta_i)^2}}{\sigma^2}\\
    &= 1 -\frac{2}{t_g \sigma^2} \int_{0}^{t_g} d\tau_1 S(\tau-\tau_1) + \frac{1}{t_g^2 \sigma^2} \int_{0}^{t_g} d\tau_1 \int_{0}^{t_g} d\tau_2 S(\tau_1-\tau_2),
\end{align}
where $S(\tau) = \overline{\eta(t + \tau)\eta(t)}$ is the auto-correlation function of the noise and $\sigma^2 = S(0)$. We can This gives us a systematic way of estimating the validity of the approximation. We can then take the time average of this quantity over a gate time to get a time dependent metric, $\epsilon$, for the validity of the constant approximation:
\begin{equation}
    \epsilon = \int_0^{t_g} d\tau' \xi(\tau').
\end{equation}
In the case of Ornstein-Uhlenbeck noise, as in the main text, we get:
\begin{equation}
    \epsilon_{\tau_c} = \frac{1}{t_g^2} (t_g^2 - 2 t_g \tau_c +2(1-e^{-t_g/\tau_c})\tau_c^2) .
\end{equation}
As expected, this vanishes in the quasistatic limit ($\tau_c/t_g \rightarrow \infty$).

\section{Comparison of our approximation with numerical simulations} \label{num_sims}
\begin{figure}[h!]
    \centering
    \includegraphics[width=1\linewidth]{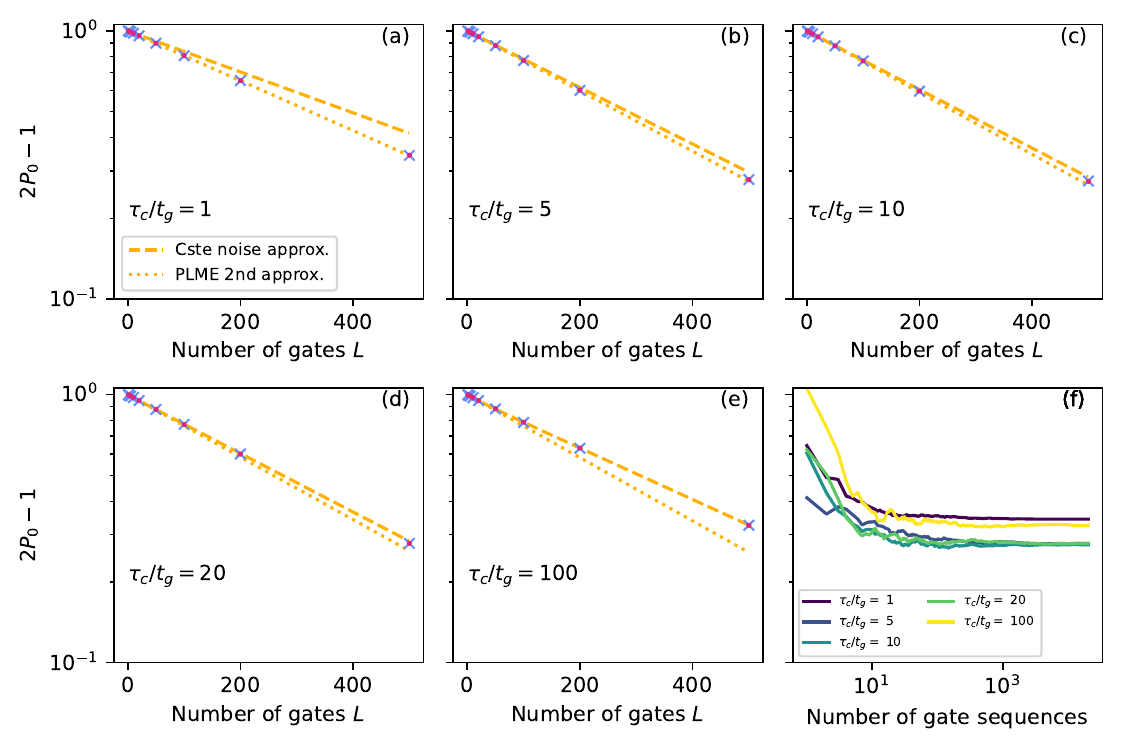}
    \caption{Validation of the constant noise approximation and of the second order PLME approximation. (a)-(e): Decay of the survival probability as a function of the number of gates. We compare noise with a correlation function $S(t)=\sigma^2 \exp(-\lvert t \rvert/\tau_c)$ with $\sigma t_g= 0.05$ and varying $\tau_c$ as specified in the plots. Our two approximations are plotted in orange and are seen to match with the stochastic averaging with and without an instantaneous first gate (blue crosses and red dots respectively). The numerical averaging was done by simulating the evolution for $2 \times 10^4$ gate sequences, each for $100$ noise realizations. The blue crosses and red dots are the total average over both the gate realizations and noise sequences with and without the initial instantaneous gate respectively. (e): Convergence of the numerical averaging. The curves represent the partial average for a subset of the gate sequences. We see that they are all converged after around 15000 gate sequences.} \label{fig:brute_force_avg}
\end{figure}

In this section, we compare the second order PLME and the constant noise approximation against brute force numerical averaging of the dynamics. To do so, we simulate the evolution of a qubit driven by the finite-duration pulses implementing the gates under the 'ZSX' gate decomposition. In order to get accurate results, we need to average over many gate sequences (here we consider 20,000 gate sequences) with 100 noise realizations each. We independently do this averaging for each sequence length $L$ as would be done in a standard RB experiment. Even by parallelizing the simulations over 60 cores, the averaging takes multiple hours for each curve depicted in Fig.~\ref{fig:brute_force_avg}, highlighting the usefulness of our analytical results. 

We compare the numerically obtained data against our two approximations for various values of the correlation time $\tau_c$. As expected, we see that for correlation times on the order of a few gate times, the second order PLME well approximates the decay of the survival probability. On the other hand, for noise with longer correlation times, we see that the constant noise approximation is valid. There is some ambiguity wether which approximation is the best in the regime where $\tau_c \approx 10t_g$, but we also see that both approximations are close to each other, meaning that either can be chosen to obtain some insight on the decay curve. 

In Fig.~\ref{fig:brute_force_avg}, we also see the decay curve in the case where the first gate is not perfect. We see that in this case, the decay is slightly different but remains qualitatively similar.

\section*{Perfect first gate approximation}
\label{sec:perfect_first_gate}

In the main text, we consider that at $t=0$, there is an initial instantaneous Clifford gate that is applied. This approximation is equivalent to saying that we can initialize the qubit perfectly in any state. While this is an approximation and is obviously not realistic in experiments, omitting it would not change the features of the decay curve qualitatively. Without this initial instantaneous gate, the only difference would be that a perfect twirl is not applied initially, meaning that errors on the first gate would not be independent of the initial state. However, this would only affect the first gate as a full twirl is applied immediately after the first gate is done and the dynamics then loses its memory of the initial states. Therefore the only consequence of the initial noiseless gate should be to slightly shift the decay of the curve due to a difference in the first gate. We see in Fig.~\ref{fig:brute_force_avg} that this difference is small (red dots vs blue crosses) and is completely negligeable in the regimes that we considered.

\section{Extension to two qubits RB}
In this section we show that our method also generalizes to the case of two qubits randomized benchmarking, where each qubit is subject to an independent, but possibly correlated noise source, $\eta_1(t)$ and $\eta_2(t)$ which couples to qubit $1$ and qubit $2$ Hermitian operators $\hat A_1$ and $\hat A_2$. The experiment consists of applying a sequence of uniformly random two qubit Clifford gates indexed by $\vec \beta$ via the Hamiltonian $\hat H_{\text{ctrl}}(t,\vec \beta$ which implements the gates via finite-pulses. The total Hamiltonian of the system in the lab frame is:
\begin{equation}
    \tilde H_{\eta_1, \eta_2}(t, \vec \beta) = \hat H_{\text{ctrl}}(t, \vec \beta) + \eta_1(t) \hat A_1 + \eta_2(t) \hat A_2.
\end{equation}
Similarly as for the one qubit case, we go in the interaction frame with respect to the noise free Hamiltonian $\hat H_{\text{ctrl}}(t, \vec \beta)$. The Hamiltonian in this frame takes the form:
\begin{align}
    \hat H_{\eta_1, \eta_2}(t,\vec \beta) &= \eta_1(t) \hat A_1(t, \vec \beta) + \eta_2(t) \hat A_2(t, \vec \beta),\\
    \hat A_i(t,\vec \beta) &= \hat g_{\beta_0}^{\dag}\hat U_0(t,\vec \beta)^{\dag} \hat A_i \hat U_0(t,\vec \beta) \hat g_{\beta_0},\\
    \hat U_0(t, \vec \beta) &= \mathcal T \exp(-i\int_0^t dt' \hat H_{\text{ctrl}}(t',\vec \beta)),
\end{align}
where we again consider that there is a zeroth instantaneous gate at the beginning of each sequence $\hat g_{\beta_0}$. In this frame, the evolution is governed by the Liouville Von-Neumann equation:
\begin{align}
    \dot{\hat \rho}_{\eta_1,\eta_2}(t,\vec \beta) &= \eta_1(t) \check{\mathcal L}_1(t, \vec \beta) + \eta_2(t) \check{\mathcal L}_2(t, \vec \beta),\\
    \check{\mathcal L}_i(t)[\cdot] &= -i [\hat g_{\beta_0}^{\dag}\hat A_i(t, \vec \beta) \hat g_{\beta_0}, \cdot].
\end{align}
Importantly, we can rewrite each superoperator $\check{\mathcal L}_i(t,\vec \beta)$ in the same way as for the single qubit gates:
\begin{equation}
    \check{\mathcal L}_i(t_\alpha,\vec \beta)=\check{\mathcal U}_0^{-1}(t_\alpha,\vec \beta) \check{\mathcal L}_{i,0} \check{\mathcal U}_0(t_\alpha,\vec \beta)
\end{equation}
where $\check{\mathcal L}_{i,0} =-i[\hat A_i,\cdot]$ is defined in the same way as before. The time evolved state is therefore:
\begin{equation}
    \hat \rho_{\eta}(t,\vec \beta) = \mathcal T \exp(\int_0^t dt' \eta_1(t') \check{\mathcal L}_1(t', \vec \beta) + \eta_2(t') \check{\mathcal L}_2(t',\vec \beta)).
\end{equation}
We can average over the gates by doing a cumulant expansion, as in the single qubit case. In doing so, we only keep the second cumulant:
\begin{align}
    &\langle (\eta_1(t_2) \check{\mathcal L}_1(t_2, \vec \beta) + \eta_2(t_2) \check{\mathcal L}_2(t_2,\vec \beta))(\eta_1(t_1) \check{\mathcal L}_1(t_1, \vec \beta) + \eta_2(t_1) \check{\mathcal L}_2(t_1,\vec \beta)) \rangle_{\vec \beta, c} = \sum_{i,j = 1,2} \eta_i(t_2)\eta_j(t_1)  \langle \check{\mathcal L}_i(t_2,\vec \beta)  \check{\mathcal L}_j(t_1,\vec \beta) \rangle_{\vec \beta}.
\end{align}
These can be computed similarly as for the single qubit case. One can also resort to the same tools as we did for the single qubit case to average over the noise (i.e the second order PLME, or the coarse-grained noise approximation), which both still apply for the same reasons in this section. One can also notice that the same argument as was used in the previous sections to justify the cumulant approximation still applies here. Whenever two times are separated by more than a full Clifford gate, the associated cumulant vanish. This can be seen as all the products of the form: $\langle \check{\mathcal L}_i(t_2,\vec \beta)  \check{\mathcal L}_j(t_1,\vec \beta) \dots \rangle_{\vec \beta}$ are completely analogous to the ones defined before, each superoperator $\check{\mathcal L}_i$ can be decomposed in the same way and the properties of the single qubit Clifford group used above also apply to the 2 qubit Clifford group.

\section{\texorpdfstring{$1/f$ noise}{Title with math}}
\begin{figure}
    \centering
    \includegraphics[width=1 \linewidth]{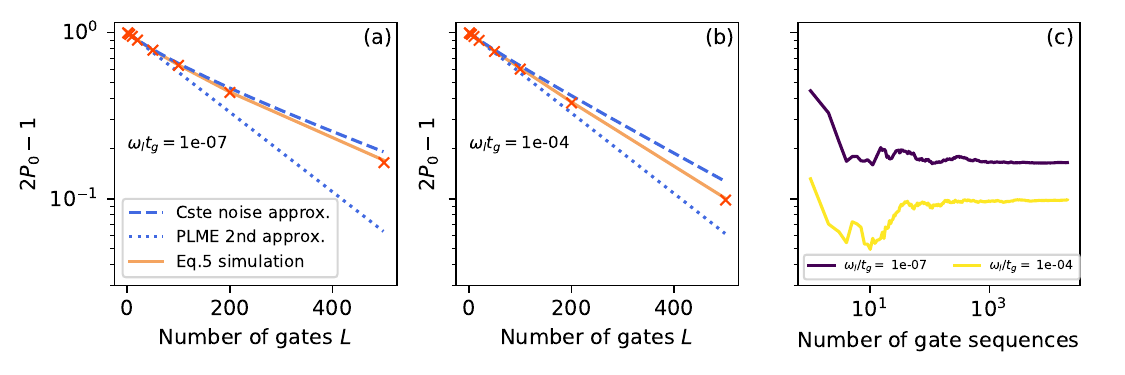}
\caption{Comparison of Eqs.~\eqref{2nd order PLME} and \eqref{surv_prob_constant_noise} in the main text with numerical stochastic averaging. In the simulations, we fixed $\lambda t_g = 0.15$ and $\omega_h t_g = 10^4$. (a)-(b): The blue curves represent the constant noise approximation and the 2nd order PLME. The red crosses are the average which results from the numerical stochastic averaging of the full dynamics which was obtained by considering $10^4$ gate sequences each with $100$ noise realizations. The orange curves are the result of the direct simulation of the second order Eq.~\eqref{gate_avg_prop}, with $10^4$ noise realizations. (c): Convergence of the stochastic evolution of the full dynamics (red crosses). The vertical axis represents the partial average (for a subset of gate sequences) of the survival probability ($2P_0-1$) as a function of the number of gate sequences in the subset. For more than $10^3$, the curve is almost flat, indicating that the numerical averaging has converged.}
    \label{fig:1overf_nums}
\end{figure}

In the main text, we mentionned that our method applies to noise models with arbitrary noise spectral density. In the section, we show that our method can easily be used to predict the decay of the survival probability in the presence of $1/f$ noise. Specifically, we consider noise with a $1/f$ spectral density with low frequecny cutoff $\omega_l$ and high frequency cutoff $\omega_h$:
\begin{equation}
    S[\omega]= 
    \begin{cases}
        \lambda^2 \frac{2\pi}{\lvert \omega \rvert} , \text{ if } \omega_l < \lvert \omega \rvert < \omega_h\\
        0, \text{ otherwise},
    \end{cases}
\end{equation}
where $\lambda$ is a parameter with units of frequency  that controls the strength of the noise. Taking the Fourier transform, we obtain the noise autocorrelation function:
\begin{equation}
    S(t) = 2 \lambda^2 (\text{Ci}(\omega_h\lvert t \rvert) - \text{Ci}(\omega_l \lvert t \rvert) ),
\end{equation}
where $\text{Ci}(t)=-\int_t^{\infty}dt' \cos(t')/t'$. Using this spectral density, we computed the survival probability decay curve using our two different approximations: the second order PLME (Eq.~\eqref{2nd order PLME}) and the constant noise approximation (Eq.~\eqref{surv_prob_constant_noise}). In Fig.~\ref{fig:1overf_nums}, we see that the validity of both approximations depend on the the low frequency cutoff $\omega_l$. This is because for a bigger $\omega_l$ cutoff, the noise spectrum is mostly formed by high frequencies while for a smaller cutoff $\omega_l$ there is an important part of the spectrum which is located at small frequencies. This means that there are important noise correlations that survive for a long time when $\omega_l$ is small, while the noise correlations mostly matter at short times for large $\omega_l$. We know from the previous sections that noise with short time correlations is well approximated by the second order PLME, while noise with long time correlations is better approximated by the constant noise approximation. This is what we see in Fig.~\ref{fig:1overf_nums}.

\end{document}